\documentclass{aa}  
\usepackage{natbib}
\bibpunct{(}{)}{;}{a}{}{,}
\usepackage{graphicx}
\begin{document}
\topmargin -1.5cm

   \title{Photometric properties of resolved and unresolved magnetic elements}

\authorrunning{Criscuoli and Rast}
\titlerunning{Photometric properties of magnetic elements}
   \author{S. Criscuoli \inst{1} and M.P. Rast \inst{2}
   	}

   \offprints{Serena Criscuoli \email{criscuoli@mporzio.astro.it}}

   \institute{  
INAF - Osservatorio astronomico di Roma, Via Frascati 33, I-00040, Monte Porzio Catone, Italia\\
\and
Laboratory for Atmospheric and Space Physics, Department of Astrophysical and Planetary Sciences, University of Colorado, Boulder CO, 80309-0391 USA\\
\\
\email{criscuoli@mporzio.astro.it, mark.rast@lasp.colorado.edu}
}

  \date{}

   \abstract
{} 
{We investigate
the photometric signature of magnetic flux tubes in the solar 
photosphere.}
{We developed two dimensional, static numerical models of isolated and clustered magnetic flux tubes. We investigated the emergent intensity profiles at different lines-of-sight for various spatial resolutions and opacity models. 
}
{We found that both geometric and photometric properties of bright magnetic features are determined not only by the physical properties 
of the tube and its
surroundings, but also by the particularities of the
observations, including the line/continuum formation height, the spatial resolution and the image analyses techniques applied. We show that some observational results presented in the literature can be interpreted by considering bright magnetic features to be clusters of smaller elements, rather than a monolithic flux tube.}
{}

   
\maketitle
   
%

\section{Introduction}
Recent observations taken in the G-band with sub-arcsecond  resolution have allowed the study of photometric \citep{berger2004,okunev2004,ishikawa2007,berger2007} and dynamic  \citep{bovelet2003,nisenson2003,rouppe2005,ishikawa2007} properties of small magnetic flux concentrations in the solar photosphere. Observations away from
disk center have revealed that bright magnetic flux concentrations are usually accompanied by a 
dark lane on their disk center side 
and by
a bright tail on the limb-ward side \citep{lites2004, hirzberger2005, berger2007}.
However, some bright points show no associated tail or dark lane, 
and some dark lanes with 
enhanced magnetic flux density show no 
associated bright point \citep{berger2007}. Somewhat 
contradictory measurements have been reported for
the center-to-limb variation (CLV) of
G-band bright point photometric contrast and size.
For example \citet{berger2007} reported maximum contrast
at a position much closer to the limb than previous observers, 
while \citet{hirzberger2005}, after analysis of extreme limb 
observations, reported no trend in contrast or size with 
position on the solar disk. Discrepant CLVs have also been 
obtained when comparing G-band with other continua measurements
\citep[e.g.][]{auffret1991,sutterlin1999,sanchez2002,okunev2004}. Together these observations have raised 
questions about the validity of viewing  
bright magnetic features as monolithic 
flux elements. In particular, it has been suggested that 
the observed CLVs of contrast and size are determined 
by properties of unresolved clusters of magnetic elements 
at the limb \citep[e.g.][]{okunev2005,berger2007}. 

The properties of small individual magnetic features have been 
intensively investigated using both two and 
three dimensional magneto-hydrodynamic numerical simulations \citep[e.g.][]{keller2004, carlsson2004,steiner2005,hasan2005,uiten2006}.
Some of the highlights of these modeling efforts include the
following.
Using both static and dynamic two-dimensional simulations, 
\citet{steiner2005} showed that the dark lane observed disk-center-ward
of bright magnetic elements is caused by the presence of 
cooler material adjacent to (both inside and outside) 
the disk-center-ward flank of the magnetic tube.  
Similar results were previously presented by 
\citet{pizzo1993a,pizzo1993b,deinzer1984, knolker1988a, knolker1988b}, who also investigated the sensitivity
of the contrast intensity profiles to the physical properties of 
the tube (such as the temperature boundary condition, 
the evacuation of the tube, the intensity of magnetic field, 
and the inhibition of convection) and its temporal evolution. 
Ultimately, the intensity contrast profiles are determined by 
the degree of evacuation of the tube and the properties 
(particularly temperature) of the atmosphere
inside and surrounding it. Fully three-dimensional 
magneto-hydrodynamic 
simulations \citep[e.g.][]{keller2004, carlsson2004} have 
demonstrated that the increase in magnetic element 
contrast with distance from disk center 
largely reflects the properties of the solar granulation 
behind the tube.  This becomes visible away from
disk center because of the reduced opacity within the tube for 
highly oblique viewing angles. 
Moreover, \citet{depontieu2006}, using both magneto-hydro-dynamic simulations and 
high spatial resolution observations, demonstrated that the
temporal evolution of the photometric properties of 
magnetic flux tubes occurs on the characteristic 
granular time scale. 

The literature addressing the properties of flux tube aggregations 
is less extensive. \citet{caccin1979} showed that the shape of the 
contrast-CLV of clusters of flux tubes is independent of the 
shape of the single underlying  model tube 
(funnels or cylinders). More recently, 
\citet{okunev2005}, following \citet{karpinsky1998},  used three-dimensional static models  in pressure equilibrium,
to show that the appearance of aggregations of flux tubes  
and their measured polarimetric signals are strongly influenced
by their position on the solar disk, filling factor and spatial resolution.  

In this paper we investigate how the photometric 
signatures of isolated and clusters of small magnetic elements  (such as the  center-to-limb variation of photometric and geometric properties) 
reflect, not only the physical properties of the tube and its
surroundings (such as temperature stratification and 
magnetic flux density),
but also the particularities of the observation
itself (such as spatial resolution, formation height of the 
wavelength of observation and image analyses technique employed).
We examine these sensitivities using
two-dimensional models of static magnetic flux tubes in 
radiative equilibrium with their surroundings. We show that results obtained by the analyses of recent high resolution observations can be interpreted assuming that observed magnetic bright features are actually aggregations of unresolved magnetic structures.   

In \S2 we describe the model. 
In \S3 we investigate thermal properties of aggregations of magnetic flux tubes. In \S4 we  present the obtained contrast profiles and discuss their dependence on observational wavelength and spatial resolution. In \S5 and \S6 we  show how the observed CLVs of photometric and geometric properties of aggregations of magnetic features depend not only on peculiarities of observations, but also on the numerical techniques employed to detect features on images. In \S7 we investigate the size dependence of contrast. 
Conclusions are presented in \S8.

\section{The Model}
We consider static models of isolated and clustered magnetic flux tubes in Radiative 
Equilibrium (RE, hereafter) with the surrounding non-magnetic plane parallel atmosphere. 
The RE is imposed with an iterative scheme similar to the one proposed by \citet{pizzo1993a} 
In brief, initial atmospheric conditions for the quiet sun and flux tube are imposed, the mean intensity everywhere in the domain is estimated and the logarithmic gradient $\nabla\equiv d\ln T /d \ln P$, where $T$ is the temperature and $P$ is the pressure, is calculated.  A new temperature field is then computed imposing RE in those layers which do not satisfy the Schwarzschild criterion. A smooth variation of temperature between convectively stable and unstable layers is imposed using a linear interpolation. Namely, in the region 70 km thick where $\nabla \simeq \nabla_{ad}$, the temperature is given by $T(x,z)=(1-\alpha)T(x,z)_{R}+\alpha T(x,z)_{C}$, where $T_{R}$ is the temperature value estimated by imposing RE, $T_{C}$ is the temperature value given by the assumed atmosphere model and $\alpha$ is a parameter which varies between 0 and 1. From the solution of the hydro-static and state (a perfect gas is assumed) equations the new pressure and density are computed respectively, while source function is estimated assuming local thermodynamic equilibrium (LTE). 
In this new atmosphere the  mean intensity and the logarithmic gradient $\nabla$ are re-evaluated. The procedure is halted when temperature relative difference among two consecutive iterations is less than 0.001$\%$. The resulting atmosphere therefore differs from the initial one only in those layers which are stable against convection.
The mean intensity is evaluated integrating over the solid angle by a Carlson \citep{carlson1963} quadrature scheme (namely, scheme A with 10 angles per octant).The intensity in each spatial direction 
is evaluated by a code \citep{criscuoli2007}\footnote{See http://dspace.uniroma2.it/dspace/advanced-search/} based on 
the short-characteristic technique \citep{kunasz1988}.
The initial non magnetic atmosphere is derived from the model of \citet{kurucz1994}; 
the opacity is assumed to follow  
a power law in pressure and temperature. 
For a complete description of the atmosphere model see
\citet{giordano2007}. 

The presence of a flux tube is approximated by shifting a portion of 
the atmosphere
downward prescribing the initial Wilson depression amplitude, 
which along with the size of the tube is a 
free parameter of the simulations. Note that, when iteration is halted, the final Wilson depression differs from the initial one.
Incoming radiation at the top of the two-dimensional
domain is set to zero while radiation at the bottom of the 
domain is estimated based on the radiative diffusion 
approximation.  Horizontal periodicity is imposed at the 
vertical boundaries. The spatial domain is sampled by a 
grid with vertical and horizontal resolutions of 7.1 km and 
3.5 km, respectively.
The vertical domain is 1130 km deep, 630 km above and 500 km 
below the optical depth unity. The horizontal extent of 
the domain is such that the ratio between the tube diameter 
(or of the cluster) and the width of the domain 
is large enough so that the presence 
of the tube does not significantly
influence 
the temperature stratification of the quiet atmosphere \citep{fabiani1992}.

The model solutions we present are thus
based on the following assumptions. The geometry we consider is 
2D, although the radiative field and quadrature employed is fully 3D. 
We expect this to reduce the effects of 
radiation channelling (see next section), although temperature 
differences between 2 and 3D calculations are estimated to be 
small \citep{uiten1998}.  
Our model is also not in horizontal pressure equilibrium, which means that the tube geometry is simplified; flaring with height is not reproduced. As a consequence, the tube contrasts near the limb 
are somewhat reduced. However, thin flux tube models suggest 
that this flaring is negligible at photospheric 
heights we consider.  
Finally, our model assumes a gray opacity.
\citet{vogler2004} showed that this assumption, by neglecting 
additional coupling of the radiation with matter 
\citep[see for instance][]{steiner1990}, leads to an under-estimation of 
the radiative illumination effect (see next section) in flux tubes, and 
therefore
to under-estimation of temperature in magnetic features. 
That author also showed that magnetic features contrast is  under-estimated 
in gray calculations. We expect the same effects in results obtained with our 
calculations.
  
The models we developed are not meant to exactly reproduce 
the results obtained by recent high resolution observations, 
but rather to show qualitatively how measurements of contrast and geometric properties of isolated and clusters of magnetic elements are affected by observational issues.

\section{Thermal properties of clusters of flux tubes}
In this section we discuss the thermal properties of clusters of magnetic flux tubes, solutions in which more than one tube 
is in RE with the surrounding atmosphere. While we have 
investigated the thermal signatures of a range of flux tube sizes
and configurations (i.e. number, strength, and separations), we 
present here only the results for aggregations of flux tubes 
with diameter 70 km, center separation of 105 km, and initial Wilson depression 150 km. 
\begin{figure}
\centering{
\includegraphics[width=7.3cm]{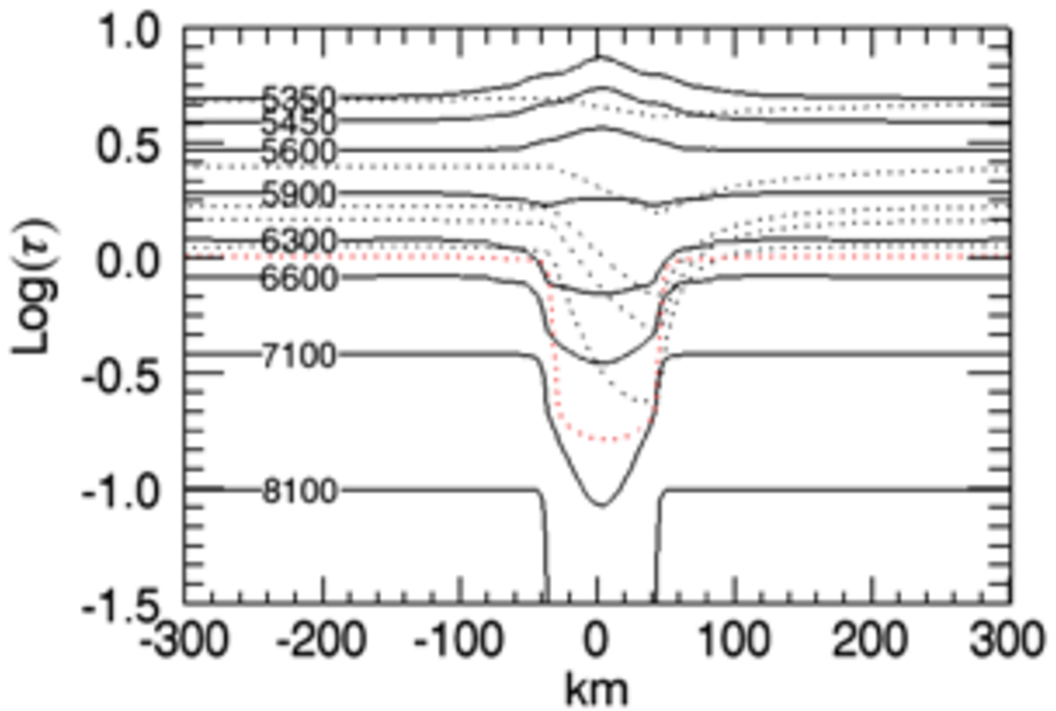}
\includegraphics[width=7.3cm]{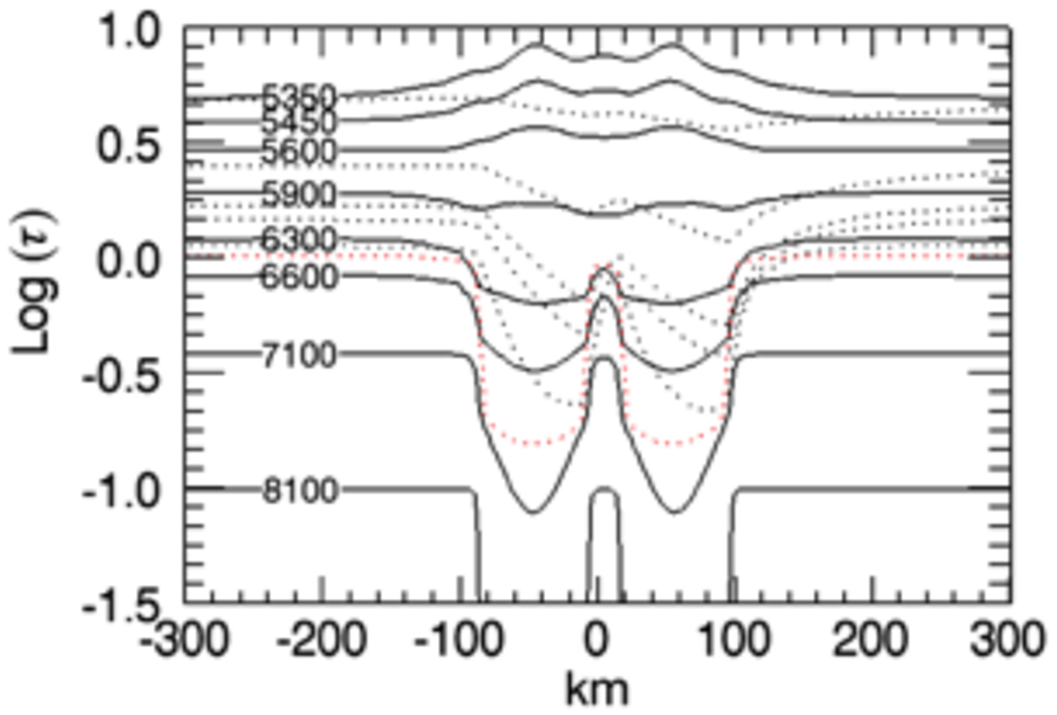}
\includegraphics[width=7.5cm]{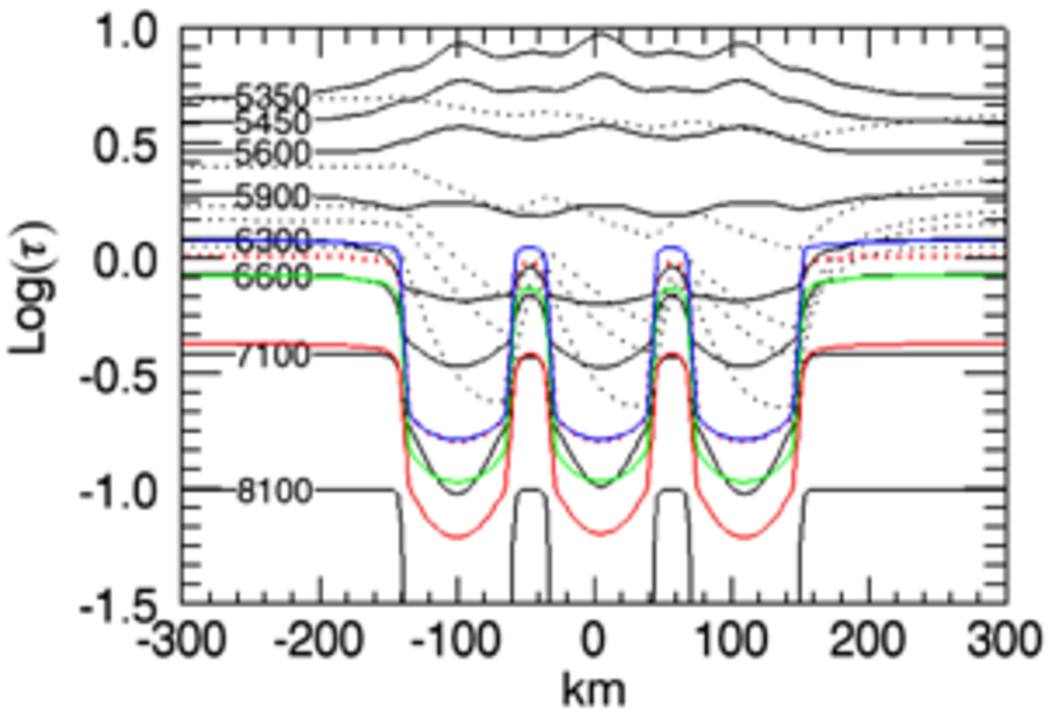}
}
\caption{Detail of the temperature field inside and around an isolated (top), two (center), three (bottom) magnetic flux tubes in RE. Diameters are 70 km and tubes axis are 105 km apart. The continuous lines represent temperature iso-contours (in Kelvin). The dotted lines represent the heights at which $\tau=1$ for various lines of sight: $\mu$ = 0.2, 0.4, 0.6, 0.7, 0.9, 1 from top to bottom. The colored lines in the bottom panel represent the $\tau=1$ depth at $\mu=1$ for various opacity models. Red dotted line: gray model. Blue continuous line: $\lambda$ = 8000 $\AA$. Green continuous line: $\lambda$ = 5000 $\AA$. Red continuous line: $\lambda$ = 16000 $\AA$.
 }
\label{profiles_temp_multi}
\end{figure}

\begin{figure}
\centering{
\includegraphics[width=7.2cm]{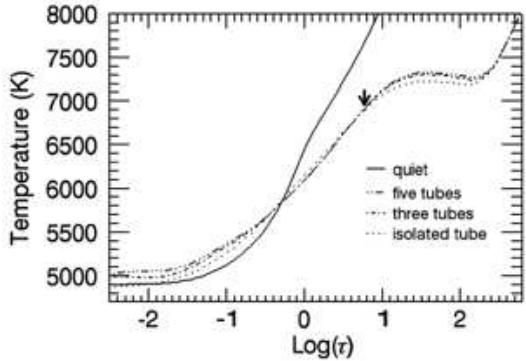}
}
\caption{Temperature along central axis versus optical depth for isolated, three and five tubes configurations in previous figures. The temperature profile of non magnetic atmosphere is given for reference. Arrow indicates optical depth unity along the axis of the isolated tube model.
 }
\label{temp_verticalprofs}
\end{figure}

Examples of the temperature fields obtained by simulations are illustrated in Fig. \ref{profiles_temp_multi}. This shows the temperature field for an isolated (top), two (middle) and three (bottom) magnetic flux tubes. 
The final Wilson depression of the tubes in RE is approximately 60 km in all the models. Using the thin flux tube approximation, the magnetic field intensity along the tubes axis estimated at optical depth $\tau=1$ in the non magnetic atmosphere is approximately 1.5 kG in all 
the configurations shown.
The temperature at the bottom of the domain is lower then the surrounding atmosphere, as imposed by boundary and initial conditions. Because of the channeling \citep{cannon1970,kneer1987,fabiani1992,hasan1999} of radiation escaping from the surrounding hotter atmosphere, 
the temperature difference between the magnetic elements and the non magnetic atmosphere decreases with decreasing optical depth. 
Eventually, in the optically thin part of the domain the temperature of the tubes and of an area in between and surrounding them exceeds the one of the non magnetic atmosphere. 
In particular, we note that between $0.5\leq\tau\leq 2$ the temperature of plasma surrounding the tubes is lower with respect to the quiet atmosphere, while it increases at shallower depths. The depth at which the channeling is effective in heating the tubes, as well as the 
area of the surrounding atmosphere affected by the presence of the tubes, 
is determined by the ratio of the tube diameter and of the horizontal optical depth value \citep{pizzo1993b,criscuoli2007}.
The heating is larger along tubes flanks, and the effect is larger for external flanks of peripheral tubes (see for instance the isotherm at 6300$^\circ$K in the three tubes case of Fig. \ref{profiles_example_multi}).

Inspection of temperature fields also reveals that the amount of heating  and cooling is a function of the number of flux tubes. This is better illustrated by the plot in Fig. \ref{temp_verticalprofs}, which 
shows the temperature variation with depth along the axis of an isolated tube and of the central tube of clusters of three and five elements. The temperature of the quiet atmosphere is given for reference. 
At heights $1<{\rm log}(\tau)<2$  heating is larger for aggregations of tubes with respect to the isolated one, but the increase of temperature is almost independent of the number of tubes in the
 cluster (three or five in this case). This is due to the fact that at these heights, because of the high value of the opacity (or of the horizontal optical depth), radiation escaping
 from the furthest tubes cannot penetrate into the central tube and contribute to its heating.
On the contrary, at $\tau<1$ heating is largely dependent on
the number of tubes. In these regions the increase of 
temperature is determined by the "illumination" from the layers at $\tau=1$ within and along the flanks of the tube \citep{pizzo1993b, knolker1988a}. The increase of the number of magnetic elements increases the width of the illuminating surface and therefore increases the temperature in the higher layers of the photosphere above the cluster.
Between $0<{\rm log}(\tau)<1$  
temperature is lower in 
clusters than near isolated tubes. There is a little dependence of this effect on the number of tubes. Note that a decrease of temperature in clusters of tubes was predicted by \citet{deinzer1984}. 

Finally, we investigated the sensitivity of the temperature field to the increase of the number of adjacent flux tubes for various magnetic field intensities (for various values of the initial Wilson depression). We found that,
when holding the number of tubes constant, the stronger the magnetic field, the 
greater the increase of temperature at $\tau<1$ and  the larger is the temperature decrease between  $0<{\rm log}(\tau)<1$. 

These results are in agreement with the ones previously obtained by \citet{fabiani1992}. Those authors investigated the dependence of temperature variations on
filling factor  by numerical 2D simulations, exploiting the horizontal periodicity of the intensity field and varying the horizontal size of the spatial domain of the simulations. 
Their models show the same temperature increase 
in the outer layers and decrease in lower layers of the atmosphere. 

\section{Contrast profiles}
\begin{figure*}
\centering{
\includegraphics[width=5.5cm, height=4.cm]{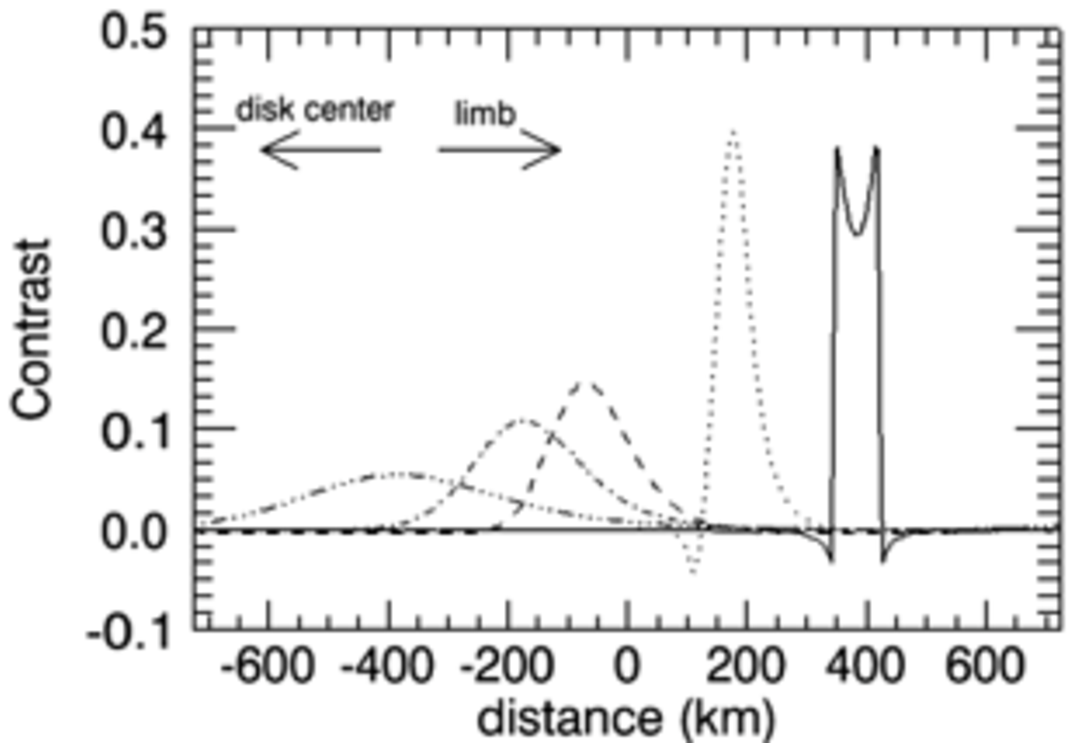}
\includegraphics[width=5.5cm, height=4.2cm]{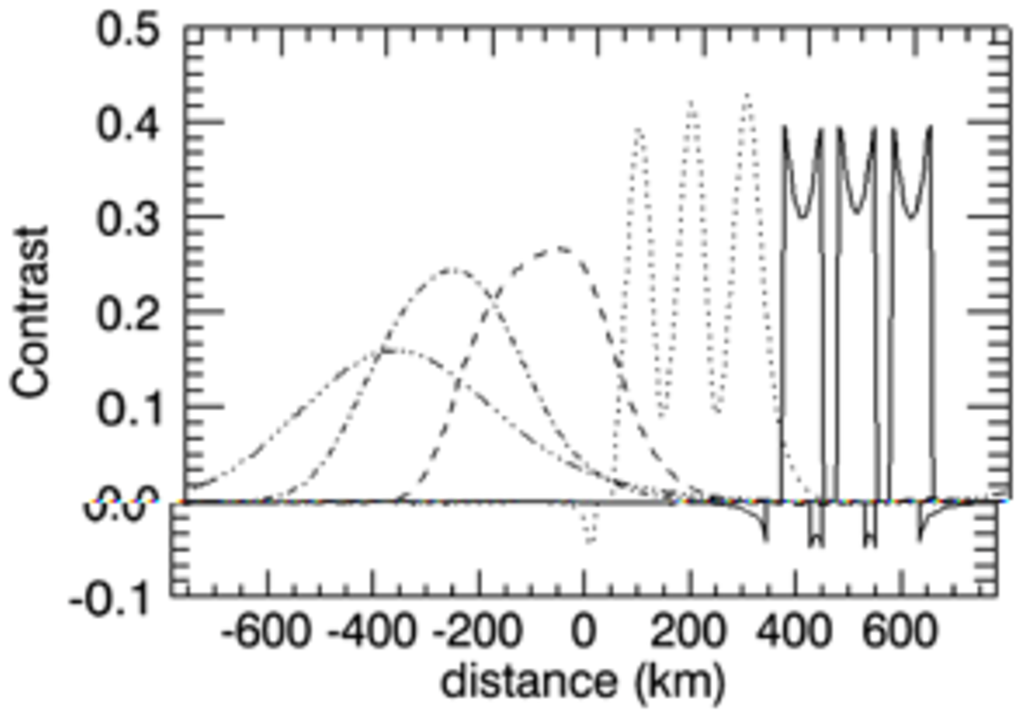}
\includegraphics[width=5.5cm, height=4.cm]{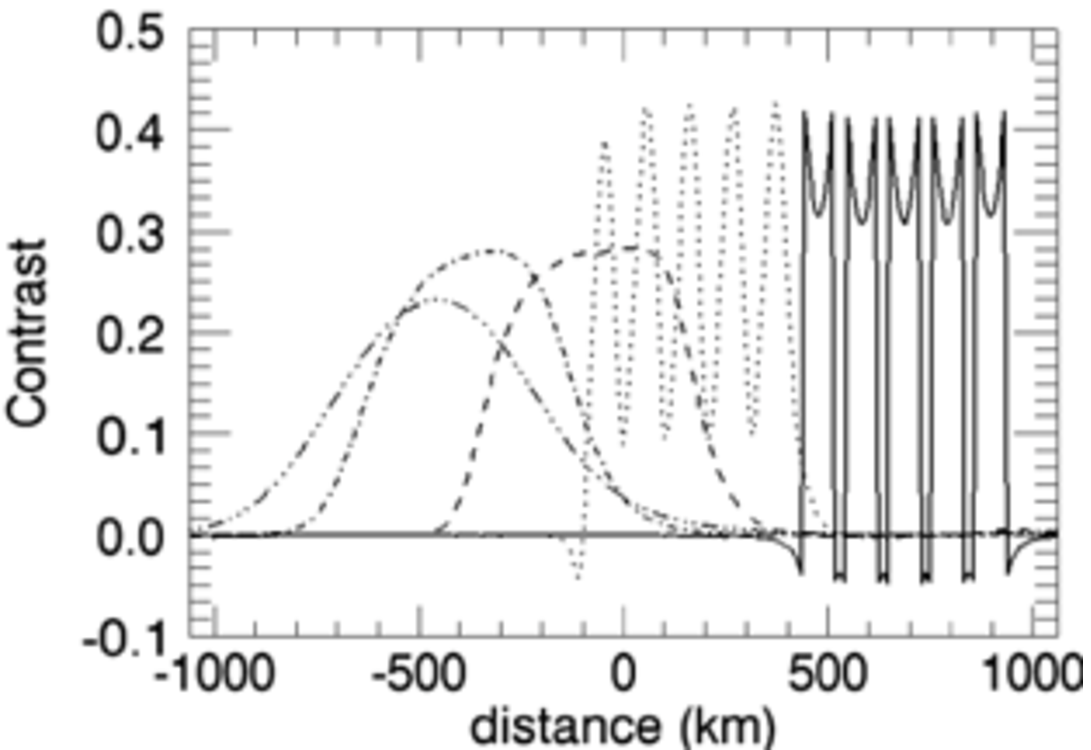}
}
\caption{Contrast profile across an isolated, three and five tubes observed at different positions on the solar disk. Tubes are 70 km wide and separated (measured from tubes axis) by 105 km. The disk center is on the left. Continuous line: $\mu=1$, dotted line: $\mu=0.9$; dashed line: $\mu=0.7$; dot-dashed line: $\mu=0.6$; dot-dot-dashed line: $\mu=0.4$}
\label{profiles_example_multi}
\end{figure*}

\begin{figure*}
\centering{
\includegraphics[width=5.5cm, height=4.cm]{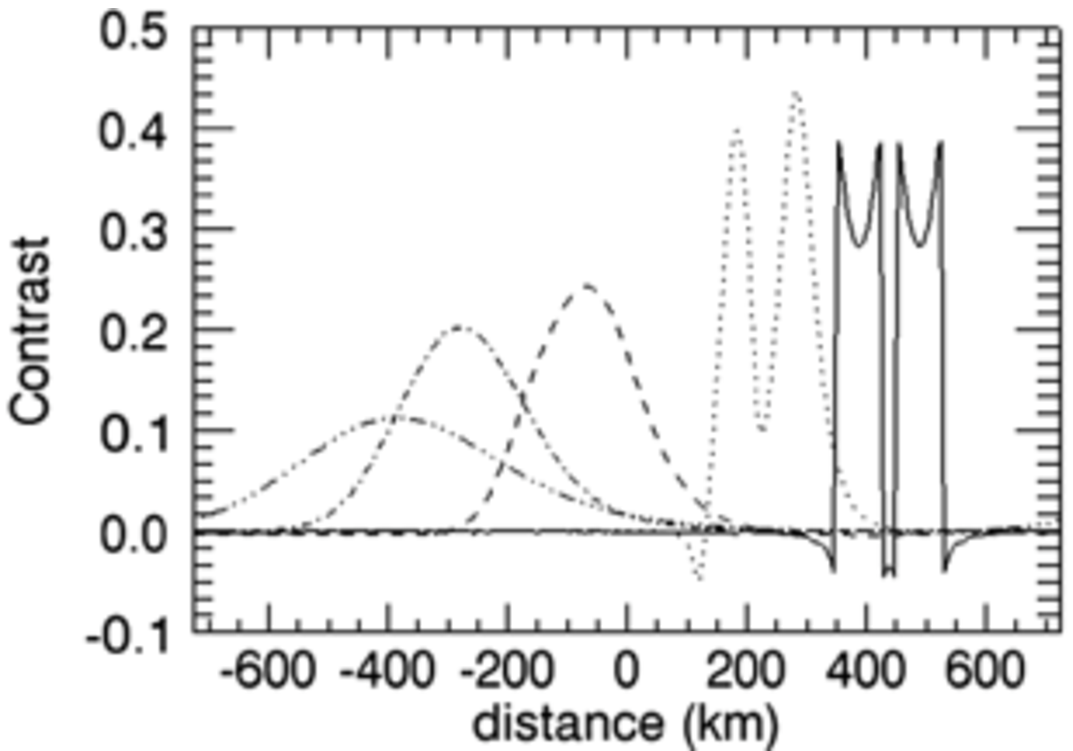}
\includegraphics[width=5.5cm, height=4.cm]{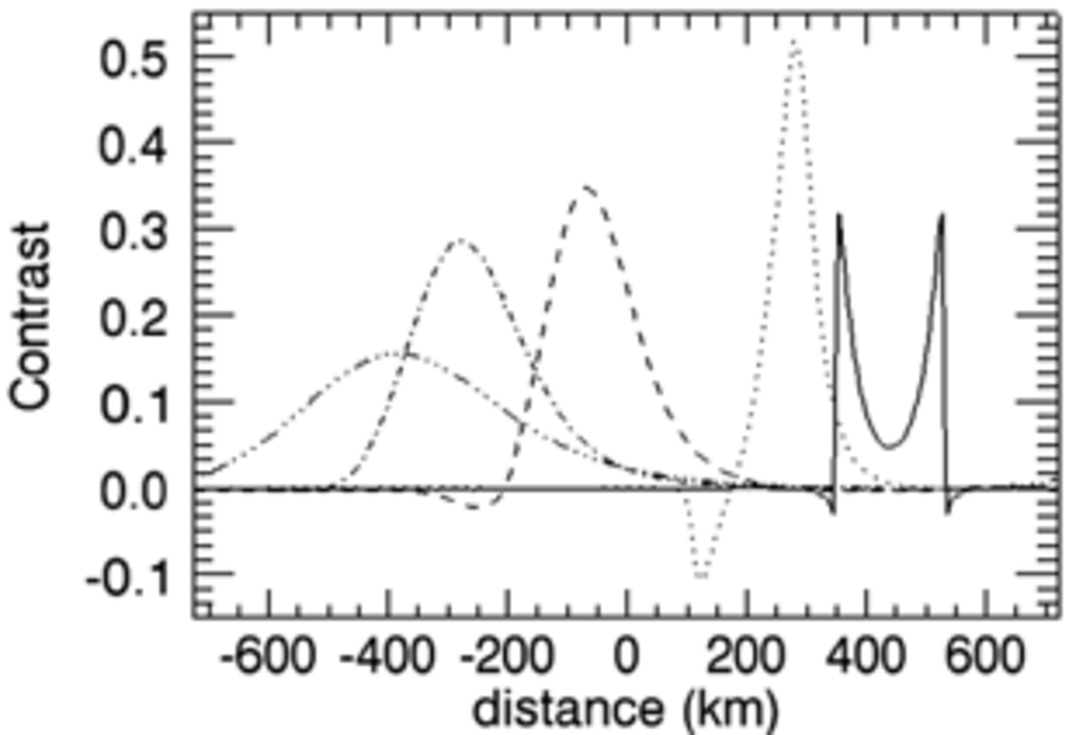}
}
\caption{Left: contrast profile across two tubes 70 km wide 105 km apart. Right: contrast profiles of an isolated flux tube 175 km wide. Legend is as in fig.\ref{profiles_example_multi}. }
\label{profiles_example_multi2}
\end{figure*}
We define the contrast to be the ratio $I(x)/ I_q-1$, where $I(x)$ is the emergent intensity value at each horizontal position \textit{x} and $I_q$ is the emergent intensity value far from the tube. We find, for various flux tube models and clusters configurations, that the contrast profiles depend on both the number of tubes and their separation distances. 
For example, Fig. \ref{profiles_example_multi} shows the contrast CLVs of 
an isolated flux tube and groups of two, three and five flux tubes of 70 km diameter and separation distance of 105 km. 
At disk center, the profiles are characterized by two positive contrast "humps" and a surrounding negative contrast "ring". The "humps" are generated by the different inclinations of the $\tau=1$ contour with respect to the isotherms as shown in Fig. \ref{profiles_temp_multi} \citep[see also][]{deinzer1984, knolker1988a,knolker1988b}. 
The "dark ring" is generated by the cool material surrounding the tube.  We 
note that the maximum contrast value at $\mu=1$ slightly increases with the number of tubes. This reflects the slight increase of temperature with the number of adjacent tubes shown in Fig. \ref{temp_verticalprofs} at the $\tau=1$ depth inside the tubes.   
Figure \ref{profiles_example_multi} also shows that the tubes are fully resolved 
only at disk center. The shallower the line of sight, the more the tubes appear as a single structure, with dimming on the disk 
center side and brightening on the 
limb side, as observed for single flux tubes. 
This is in agreement with the findings of \citet{okunev2005}, who showed that at the limb aggregations of simulated faculae appear as chains of small bright points.
For lines of sight at which the tubes are still distinguishable, the ones which are closer to the limb appear brighter than the tubes closer to the disk. 
This is due to the fact that the radiation emitted at the limb side crosses more than one tube and is therefore less attenuated, or, in other words, the $\tau=1$ contour penetrates deeper into the atmosphere, thus sampling higher temperature layers, as also shown in Fig. \ref{profiles_temp_multi}.
For similar reasons, the values of contrast at inclined lines of sight increase with the number of tubes in the cluster. Since for a given inclination and cluster configuration there is a finite number of tubes crossed by each ray, contrast increases with the 
number of tubes until this maximum number is reached. This maximum number increases with the inclination of the line of sight. 

Figure \ref{profiles_example_multi2} displays solutions for isolated and double flux tubes which occupy the same physical space: 175 km. At disk center the contrast of  the isolated large flux tube is lower with respect to the two small tubes case, while the contrary is observed off disk center. 
The opacity of the larger single flux tube is smaller than that of the two small tubes, and the $\tau=1$ contour penetrates deeper into the hotter non magnetic atmosphere. The consequent 
contrast profiles at different positions on the solar disk are quite different for the two configurations up to approximately $\mu=0.6$. In particular, we note that the disk-center-ward
dark lane is deeper and larger for the larger isolated flux tube 
than it is for the cluster. 
At smaller values of $\mu$ the contrast profiles of the two configurations 
are quite similar.

 \subsection{Dependence on opacity model and spatial resolution}
We have computed the contrast profiles of isolated and clusters of flux tubes 
after varying the model opacity. 
This has allowed us to investigate the dependence of observed contrast profiles on the wavelength used for the observations.
Continua opacities in three different wavelengths, 5000, 8000 and 16000 \AA, were computed by using classical formula for continua processes and scattering of principal elements \citep [see for instance][]{stix2002}. No blanketing line effects were considered. Abundances were taken from \citet {grevesse1993}. 

\begin{figure}
\centering{

\includegraphics[width=7cm]{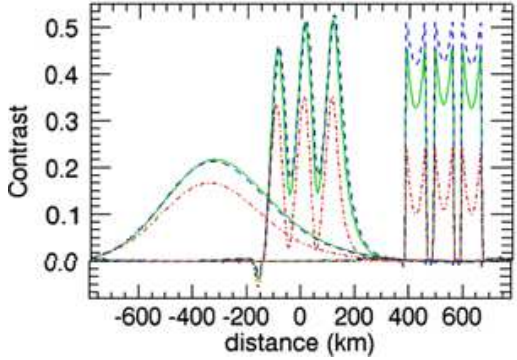}
}
\caption{Contrast profile across three tubes 70 km wide 105 km apart for various lines of sight and three different wavelengths. Blue, dot-dashed line: $\lambda$ = 8000 \r{A}. Green, continuous line: $\lambda$ = 5000 \r{A}. Red, dashed line: $\lambda$ = 16000 \r{A}. The disk center is on the left. From right to left: $\mu=1$, $\mu=0.9$ and $\mu=0.4$. }
\label{D70_d35_lambdas}
\end{figure}

We find that, because of the heating of the upper layers of the flux tubes, 
the contrast increases with the height of the  $\tau=1$ depth at
the wavelength being considered. 
The increase of contrast is larger at disk center, where larger variations of the Wilson depression occur, as shown for instance for the cluster of three tubes in Fig. \ref{profiles_temp_multi}. Figure \ref {D70_d35_lambdas} shows the corresponding contrast profiles at various lines of sight. 
We note that the negative contrast area surrounding the tubes at disk center, which is present for the corresponding gray atmosphere model illustrated  in Fig. \ref{profiles_example_multi}, is significantly reduced at the three wavelengths investigated. In fact, as shown in Fig.\ref{profiles_temp_multi}, in the grey model the $\tau=1$ line forms at a height at which the temperature of the plasma surrounding  the tubes is lower with respect to the one of the non magnetic atmosphere. At wavelengths which form at larger and shallower depths the temperature differences between the area surrounding the tube and the quiet atmosphere are reduced and the dark ring is not observed.
Variation with wavelength  largely affects the contrast values 
for vertical lines of sight, as 
can be seen by comparison of contrast profiles illustrated in Fig. \ref{D70_d35_lambdas}. In particular, the absolute contrast of dark lanes increases at wavelengths which sample the deepest layers (i.e. at optical grey depths larger than one), while they gradually disappear at wavelengths which sample shallower layers of the atmosphere (i.e. at optical grey depths smaller than one). Finally, the size of the bright area of the profiles (see next section) observed off disk center increases with the height of unity optical depth.

\begin{figure*}
\centering{
\includegraphics[width=6cm]{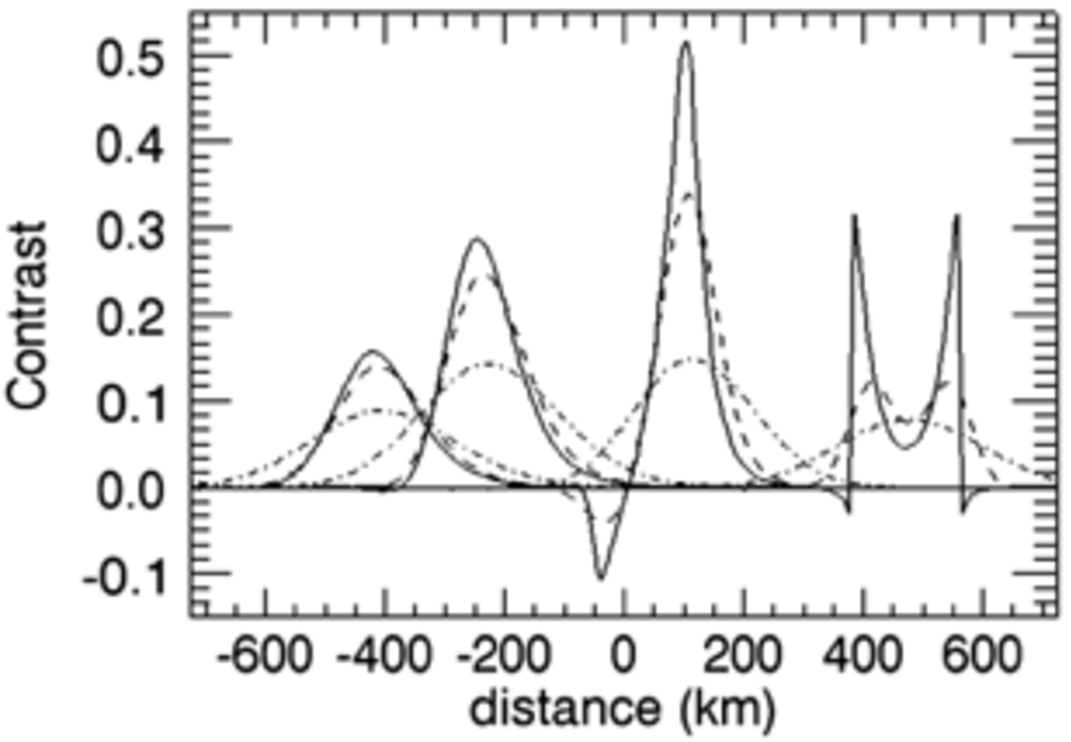}
\includegraphics[width=6cm]{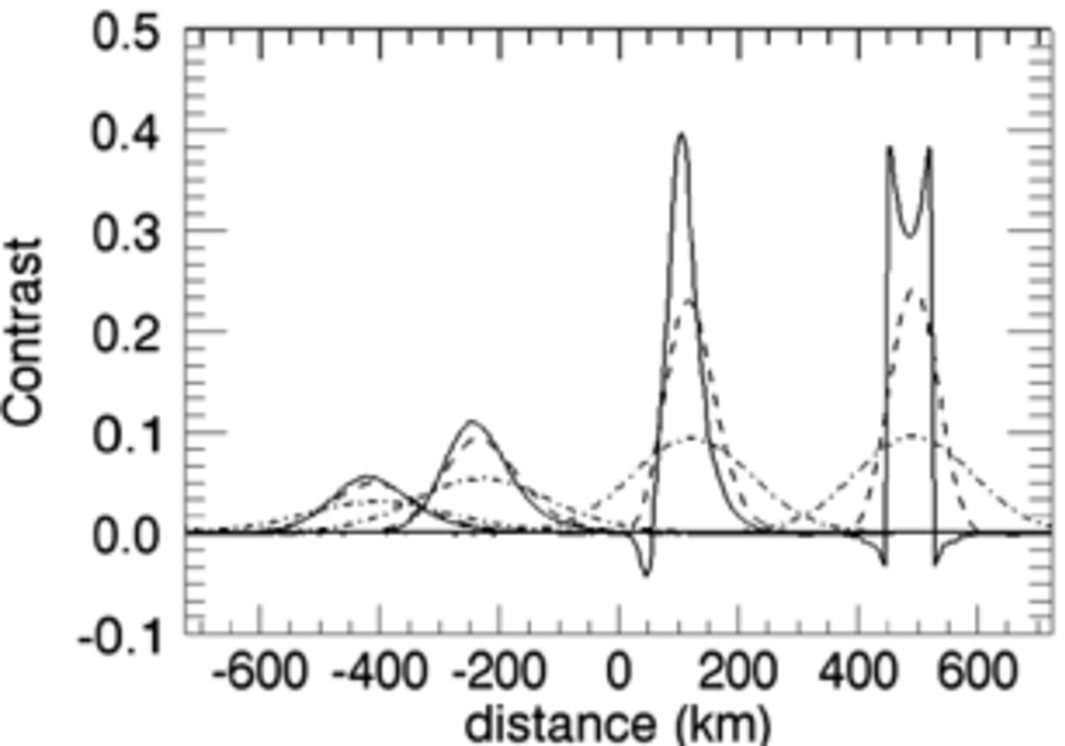}\\
\includegraphics[width=6cm]{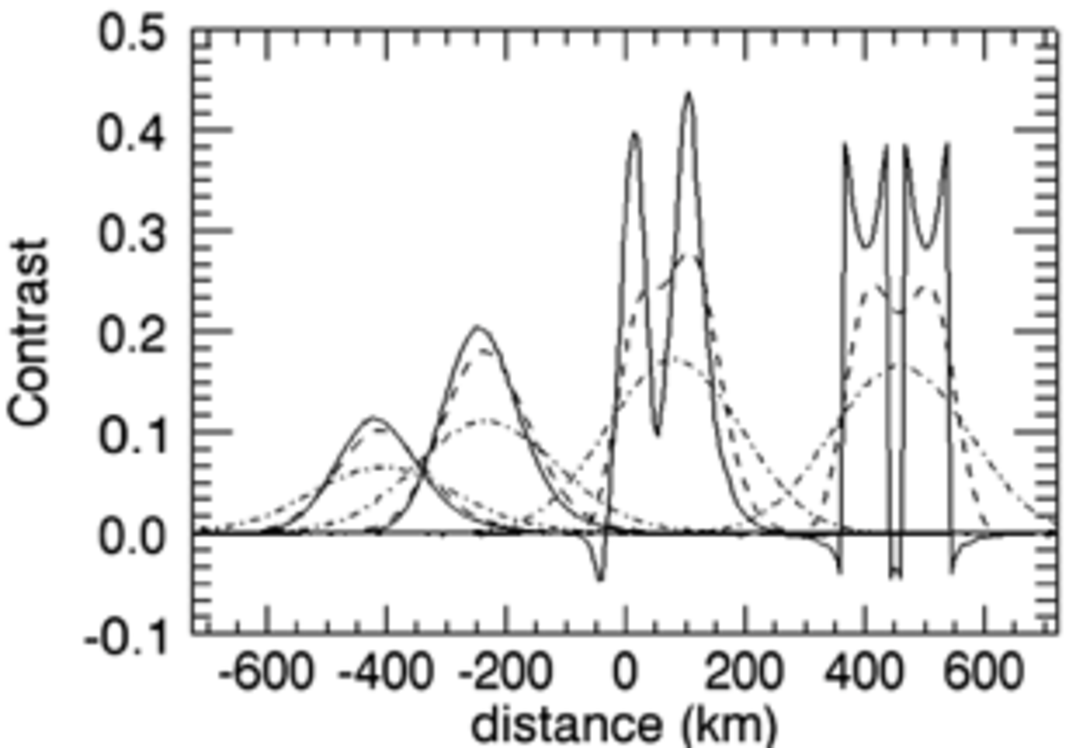}
\includegraphics[width=6cm]{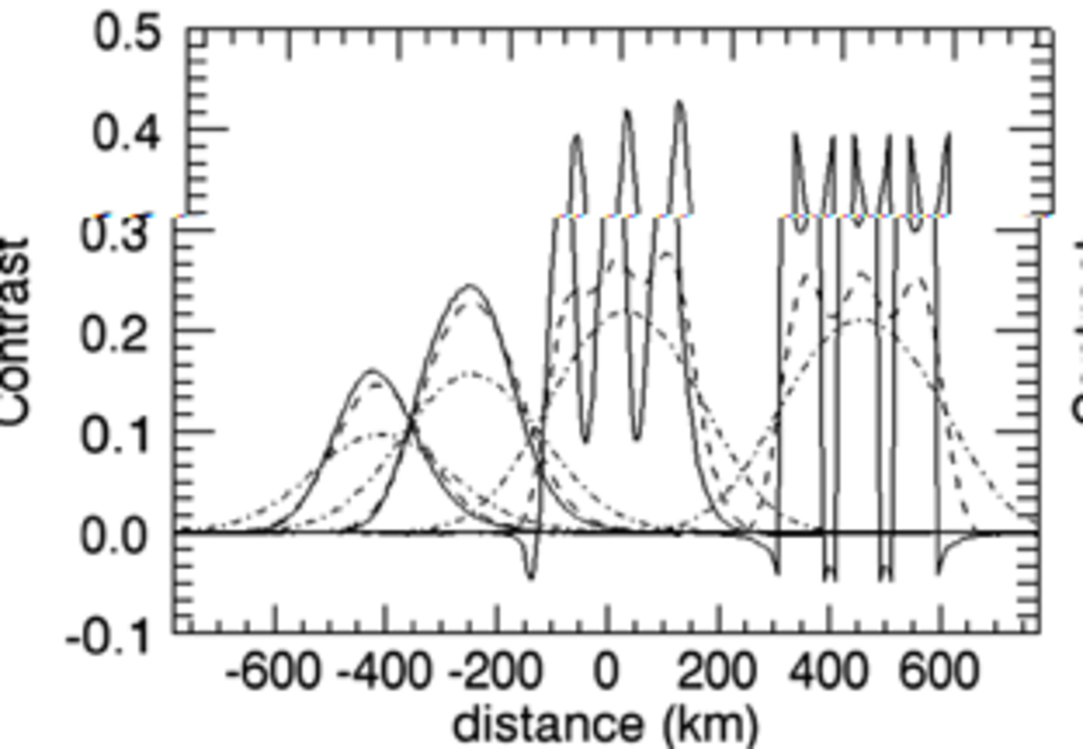}
}
\caption{Contrast profiles across flux tubes projected onto the plane of the sky for different resolutions and positions on the solar disk. Top Left: single flux tube 175 km wide. Top right: single flux tube 70 km wide. Bottom left: two flux tubes 70 km wide 105 km apart. Bottom right: three flux tubes 70 km wide 105 km apart. Continuous line: resolution of simulations. Dashed line: resolution 0.1". Dot-dashed line: resolution 0.3". In each panel, from right to left: $\mu=1$, $\mu=0.9$, $\mu=0.6$, and $\mu=0.4$.  
}
\label{profiles_example_res}
\end{figure*}

In order to investigate the effects of spatial resolution on observations,  simulated contrast profiles have been projected onto the plane of the sky, that is the contrast profiles have been re-sized by the factor $\mu$, and convolved with Gaussian functions of different Full Width Half Maxima. 
We found that projection reduces the maximum  contrast and the amplitude of the limb ward tail, thus producing  smaller and more symmetric profiles. This is illustrated for instance by comparison of plots in 
Fig. \ref{profiles_example_res} with those of 
Fig. \ref{profiles_example_multi}.  These show the contrast profiles of various flux tubes configurations after and before projection, respectively.
Reduction in spatial resolution causes decrease of maximum contrast, broadening of profiles and  smearing of small scale features, such as the dark lanes and the double "humps". These effects are larger for tubes observed at vertical lines of sight, and smaller for shallower lines of sight, where 
the intrinsic profiles are already blended by line of sight effects (Fig. \ref{profiles_example_res}). In particular, due to the presence of dark features, at vertical lines of sight the decrease of contrast values due to the reduction of spatial resolution is larger than at shallower lines of sights. As will be discussed in next section, these variations change the shape of 
the CLV. It is also interesting to note that in the models shown in Fig. \ref{profiles_example_res} with resolution 0.1", the dark lane is present in the case of the isolated large flux tube, but is not present for clusters of small magnetic elements. Inspection of Fig. \ref{profiles_example_res} reveals that the effects of reduction of resolution also depend 
on the number of flux tubes in  a cluster. For instance, maximum contrast observed at disk center is approximately the same for the isolated, two and three tubes clusters at the resolution of the simulations. With the decrease of resolution, instead, a maximum contrast proportional to the flux tube number is observed for the three configurations. 
\begin{figure}
\centering{
\includegraphics[width=7cm]{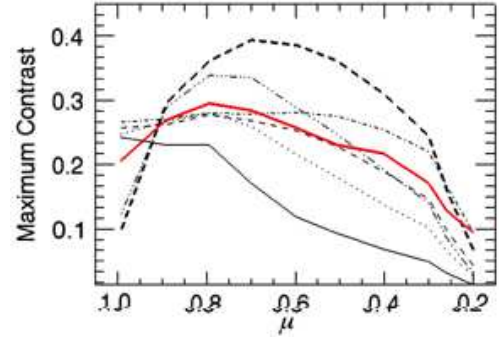}
}
\caption{CLV of maximum contrast for various flux tubes models and clusters at resolution 0.1". Continuous: single flux tube 70 km wide. Dotted: two flux tubes 70 km wide 105 km apart. Dashed: three flux tubes 70 km wide 105 km apart. Dot-dashed: five flux tubes 70 km wide 105 km apart. Dot-dot-dashed: single flux tube 175 km wide. Dashed-thick: single flux tube 280 km wide. Initial Wilson depression is 150 km for all the models. Continuous red thick line: CLV of average maximum contrasts of all flux tubes configurations computed imposing a contrast threshold of 0.08.}
\label{CLV_contrast_configs}
\end{figure}

\begin{figure}
\centering{
\includegraphics[width=7cm]{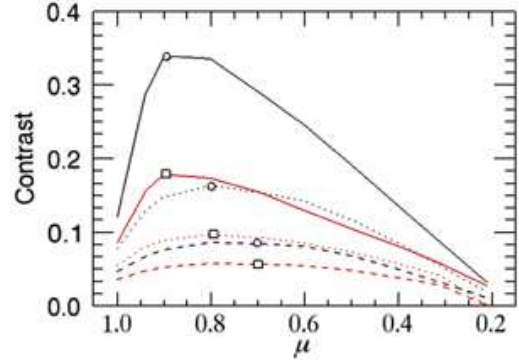}
}
\caption{CLV of maximum (black) and mean (red) contrast of a fluxtube 175 km wide. Spatial resolution is 0.1" (continuous), 0.3" (dotted) and 0.6" (dashed). Circles and squares indicate the maxima of the curves.
}
\label{CLV_risoluzione}
\end{figure}
\section{CLV of maximum and mean contrast}
We have estimated the CLV of maximum and mean contrasts of various flux tubes models and configurations. The mean contrast 
is defined as the average of contrast values larger than a specified threshold. 
Since for aggregations of flux tubes several distinct features can be selected in this way, we considered the average value of the contrast of each feature.   
Variations of the CLVs of mean contrast with variation of the threshold value were also investigated. Since the application of an intensity or contrast  threshold is usually employed to detect features on images, this study 
gives an indication of the dependence of the measured contrast CLV 
on the identification method adopted.
\begin{figure}
\centering{
\includegraphics[width=7cm]{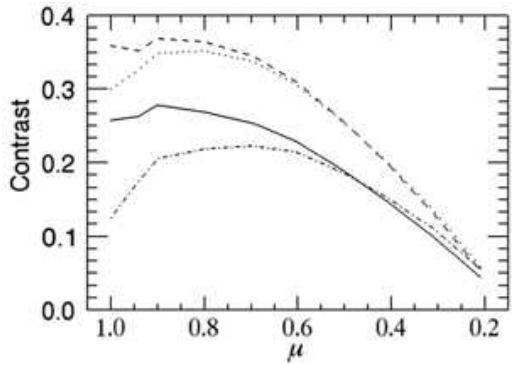}
}
\caption{CLV of maximum contrast of a cluster of three flux tubes 70 km wide 105 km apart observed with spatial resolution 0.1" in various continua. Continuous line: grey atmosphere. Dashed line: $\lambda$ = 8000 \r{A}. Dotted line: $\lambda$ = 5000 \r{A}. Dot-dashed line: $\lambda$ = 16000 \r{A}.}
\label{CLV_maxcontr_lambdas}
\end{figure}

We found that the CLV varies significantly with the number of flux tubes in a cluster. This is illustrated by Fig. \ref{CLV_contrast_configs}, in which the CLV of maximum contrast, in the case of spatial resolution 0.1", is plotted for various models of isolated and clustered flux tubes. The plot shows that the curves flatten and the peak moves toward the limb as the number
of tubes increases. The shift of the peak with 
an increase in filling factor was also found in 3-D simulations by \citet{okunev2005}. In particular, it is interesting to note 
that the curve in  Fig. \ref{CLV_contrast_configs} obtained for a cluster of five elements, resembles qualitatively and quantitatively the CLV obtained by those authors  for a cluster of elements of radius 100 km and filling factor 0.2.
The plot shows clearly that contrast of large monolithic structures (isolated flux tubes) is lower than the contrast of smaller structures at disk center, while the opposite is observed off disk center. A sharp decrease from the center to the limb is observed for the isolated smallest structures, while smoother CLVs are observed for clusters of small elements. This finding indicates that at the limb clusters and large tubes are more likely to be observed rather than isolated small tubes and therefore weigh more 
in the estimation of the CLV of contrast derived by observations. 

To investigate this effect we computed the average contrast at each position on the solar disk of the various models 
shown in the figure. To mimic selection effects associated with  the application of intensity threshold criteria on images, average has been computed discarding contrast values smaller than 0.08. The obtained CLV is represented by the red line in the plot of Fig. \ref{CLV_contrast_configs}. The resulting CLV is flatter, especially at the limb, than
the CLV of isolated tubes. \citet{spruit1976} investigated the CLV of maximum contrast obtained connecting the maxima of the CLVs of various tubes models. The author considered four classes of models and, in agreement with our findings,  showed that for three classes of models the curves so obtained were flatter than curves of individual tubes models, although the shapes of the curves were very different from the ones we have obtained. 

As already noted for the contrast profiles, we found that the maxima of both maximum and average contrasts shift toward lower values of $\mu$ with the decrease of resolution for models in which dark lanes are large and dim. This  effect is reduced in the case of aggregations of small flux tubes, for which dark lanes are less pronounced. 
 Figure \ref{CLV_risoluzione} shows for instance the CLV of maximum (black lines) and mean (red lines) contrast in the case of an isolated tube 175 km wide with initial Wilson depression 150 km. Different line-styles represent different spatial resolutions. Squares and circles indicate the maxima of the curves, which shift toward the limb at the decrease of resolution. Here we define as mean contrast the average of contrast values larger than 0.02.  The
plot also shows that, for a given spatial resolution, maximum 
and mean contrast values are correlated, i.e. the shapes of the curves are similar although the contrast values are different. We also notice that with the decrease of resolution the curves flatten  and the decrease of contrast toward the limb is less abrupt. The dependence of the contrast-CLV 
on spatial resolution was  previously reported by \citet{okunev2005}, although a systematic shift of the peak with the decrease of resolution was not observed. This is most likely due to the lack of negative contrast area (dark lanes) in their simulations.

We also investigated the dependence of CLV of the mean contrast on the contrast threshold value applied. We found a 
change in the contrast values measured, but not substantial variation of the CLV shapes. 
Finally, we found that the contrast-CLV is significantly
dependent also on the opacity model, that is on the wavelength of observations. Figure \ref{CLV_maxcontr_lambdas} plots the CLV of maximum contrast of a cluster of three flux tubes 70 km wide and 105 km apart for the opacity models 
previously discussed. The largest changes in
contrast occur at disk center (contrast decreases at wavelengths which form in the deepest layers), while at the extreme limb, 
contrast values are nearly independent of observation
wavelength. As a consequence, 
the position of the peak of the CLV shifts from $\mu\approx 1$ 
at wavelengths which form in the
highest layers to lower values of $\mu$ at wavelengths which 
form in the deepest layers.

\begin{figure}
\centering{
\includegraphics[width=7cm]{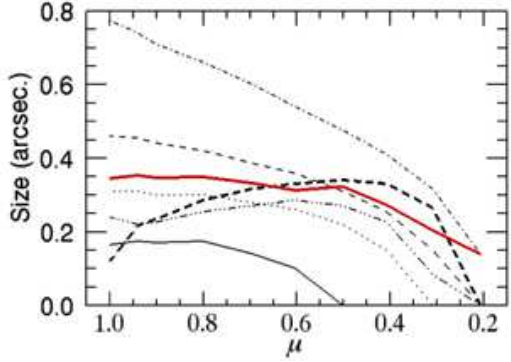}
}
\caption{CLV of size of different flux tubes models and clusters for contrast threshold value 0.08. Legend as in Fig. \ref{CLV_contrast_configs}.  }
\label{CLV_aree_configs}
\end{figure}
  
\begin{figure}
\centering{
\includegraphics[width=7cm]{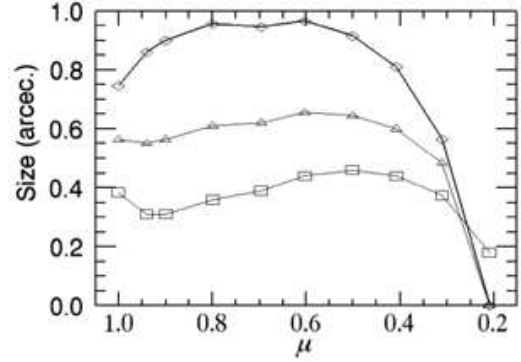}
}
\caption{CLV of size of a tube 175 km wide for various spatial resolutions. Square: 0.1". Triangle: 0.3". Diamond: 0.6". Contrast threshold value is 0.02. }
\label{CLV_aree}
\end{figure}
\section{Size and asymmetry}
We have investigated the geometric properties of isolated and clustered flux tubes. This analysis was performed on contrast profiles projected onto the plane of the sky and convolved with gaussian functions of different widths as explained in the previous paragraph.

We define the size of magnetic features as the largest distance between the points over which the contrast exceeds a given threshold.  We find that the shape of the size-CLV 
is very dependent on the flux tube models investigated, as illustrated in Fig. \ref{CLV_aree_configs}. In general, CLVs of  isolated or clusters of tubes which present marked double humps and dark lanes, show a minimum at approximately $\mu=0.9$ and a peak between $\mu=0.6$ and $\mu=0.4$. An example is given by dot-dot-dashed line in Fig. \ref{CLV_aree_configs},  which represents the size-CLV of a tube 175 km wide and initial 
Wilson depression 150 km. The decrease of size from disk center to $\mu\approx0.9$ is given by the smooth disappearance of the disk-centerward side wall and by the appearance of the dark lane, whose presence reduces the size of the "bright" area. The rise of size at intermediate lines of sights is due to the increase 
in width of that portion of the $\tau=1$ contour, which penetrates into the non magnetic atmosphere, as shown for instance in Fig. \ref{profiles_example_multi}. The steep decrease at the extreme limb is due to both foreshortening, and a decrease in the size of the region with contrast values larger than the threshold applied. A CLV of this type was found by \citet{steiner2005} for a static flux tube 170 km wide and Wilson depression 150 km.  For models which do not present marked dark lanes or humps at disk center we found monotonic variations of size with position on the solar disk, as illustrated by most of the curves in Fig. \ref{CLV_aree_configs}. The other exception in that figure is the CLV of the 280 km 
wide tube, for which an increase from the center up to $\mu=0.4$ is observed. 
This is due to the fact that the contrast profile at disk center of this model is characterized by two positive contrast humps separated by a dark lane. According to our definition, the two humps are detected as two distinct features whose average size is the estimate of the size of the tube. This definition underestimates the size of the tube at disk center, thus generating the shape of the CLV illustrated. Finally, the red line is the average value of the sizes obtained by the models in the figure for each line of sight investigated. As for the contrast, the average size-CLV is flatter than the single models, especially at the limb.

\begin{figure}
\centering{
\includegraphics[width=7cm]{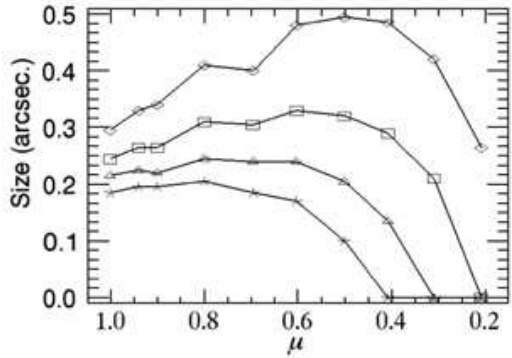}
}
\caption{CLV of size of a tube 70 km wide at resolution 0.1" for different contrst threshold values. Diamonds: 0.005. Square: 0.02. Triangle:0.04. Asterisks:0.06.  }
\label{CLV_aree_thr}
\end{figure}

The size-CLV curves also assume quite different shapes with variation 
in resolution. In general, the peak of the curve shifts toward disk center and the decrease of size toward the limb is steeper at lower spatial resolutions (Fig. \ref{CLV_aree}). 
Moreover, while a decrease in the intensity
threshold used to define structures 
corresponds an increase of the measured size, the CLV profiles remain
similar to the ones obtained at lower resolution. An example of these effects is given in Fig. \ref{CLV_aree_thr}, which shows the size-CLV 
of a tube 70 km wide for four values of threshold contrast. It is worth noting
that for high contrast threshold values or at poor spatial resolutions, 
small faint features are not detected.

\begin{figure}
\centering{
\includegraphics[width=7cm]{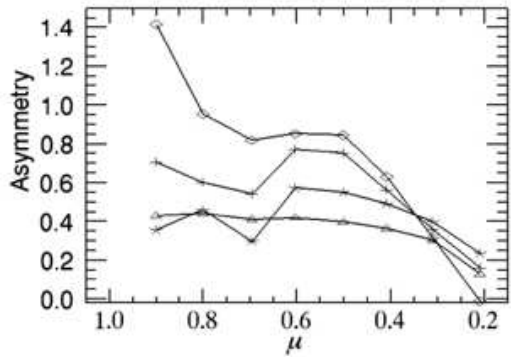}
}
\caption{CLV of skewness at various spatial resolutions for a flux tube 175 km wide. Diamonds: original. Plus: r = 0.1". Asterisks: r = 0.3". Triangles: r = 0.6".  }
\label{CLV_asym}
\end{figure}
As pointed out previously, projection onto the plane of the sky and  reduction of resolution modify the shape of contrast profiles. To quantify these variations we analyzed the skewness of the profiles and defined as "asymmetry" the skewness of distribution of contrast  values larger than a certain threshold. An example of the CLV of the asymmetries for threshold contrast value 0.001 and various spatial resolutions is illustrated in Fig. \ref{CLV_asym} in the case of a flux tube 175 km wide. We found that, in general, the skewness value is quite low, thus indicating that the contrast profiles are, overall, quite symmetric. The 
observed asymmetries are larger close to disk center and decrease toward the limb for the best spatial resolution investigated. Nevertheless, the decrease of resolution mostly affects  vertical lines of sight, therefore for intermediate resolutions the asymmetries have a peak at approximately $0.4<\mu<0.6$. At the lowest
resolution investigated the asymmetries decrease again monotonically from disk center to the limb. 
 It is also worth noticing that at $\mu<0.4$ profiles are 
nearly symmetric. This is not in disagreement with findings of previous authors, who 
reported an increase of the limb ward tail with the inclination of the line of sight \citep[e.g][]{steiner2005,pizzo1993b,deinzer1984}, 
since here we are considering profiles project onto the plane of the sky and, as previously shown, projection mostly affects the extreme limb. Moreover, this is in agreement with \citet{berger2007}, who found that experimental average contrast profiles at $\mu=0.4$ are more symmetric than average profiles at $\mu=0.6$. We also found that the value of the asymmetries defined in this way increases with the size of isolated flux tubes and with the number of tubes in a cluster.
Finally, it is also important to notice that the skewness poorly describes the asymmetries of distributions which have multiple peaks and is therefore not a good indicator of the asymmetries for profiles of isolated large flux tubes at $\mu>0.9$, or in the cases for which the profiles of clusters of tubes show multi-peaks.  

\section{Contrast-size relation}
Scatter plots of measured contrast against magnetic feature size presented in the literature usually show a large dispersion. Fits to these points have shown no trend \citep{berger1995,wiehr2004,sanchez2008} or a slight increase \citep{bovelet2003,hirzberger2005,berger2007} of contrast with magnetic feature size. In particular, \citet{berger2007} showed that the contrast-size dependence of magnetic features for various positions on the solar disk can be fitted by parallel straight lines. 
A closer inspection of their Fig. 5 reveals that, especially for off-disk center features, the increase of contrast is sharper at the smallest areas, while no trend is observed at the largest areas. 
This is in agreement with \citet{bovelet2003} and \citet{hirzberger2005} and also qualitatively with results obtained by analyses of full-disk images \citep[see][and references therein]{ermolli2007}.   
\begin{figure}
\centering{
\includegraphics[width=7cm]{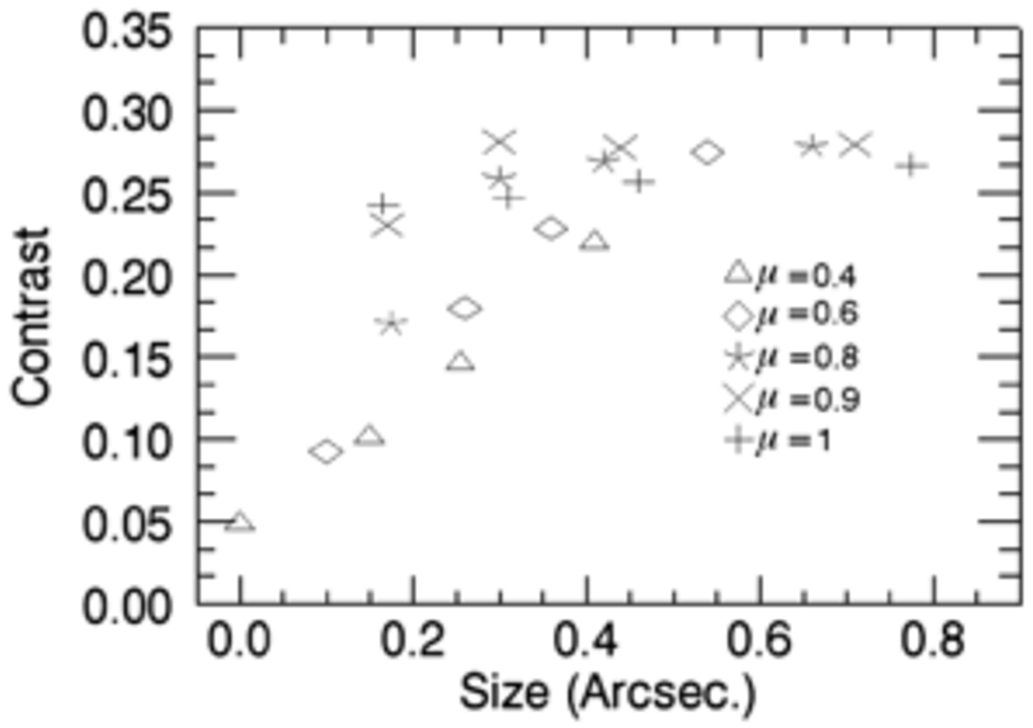}
\includegraphics[width=7cm]{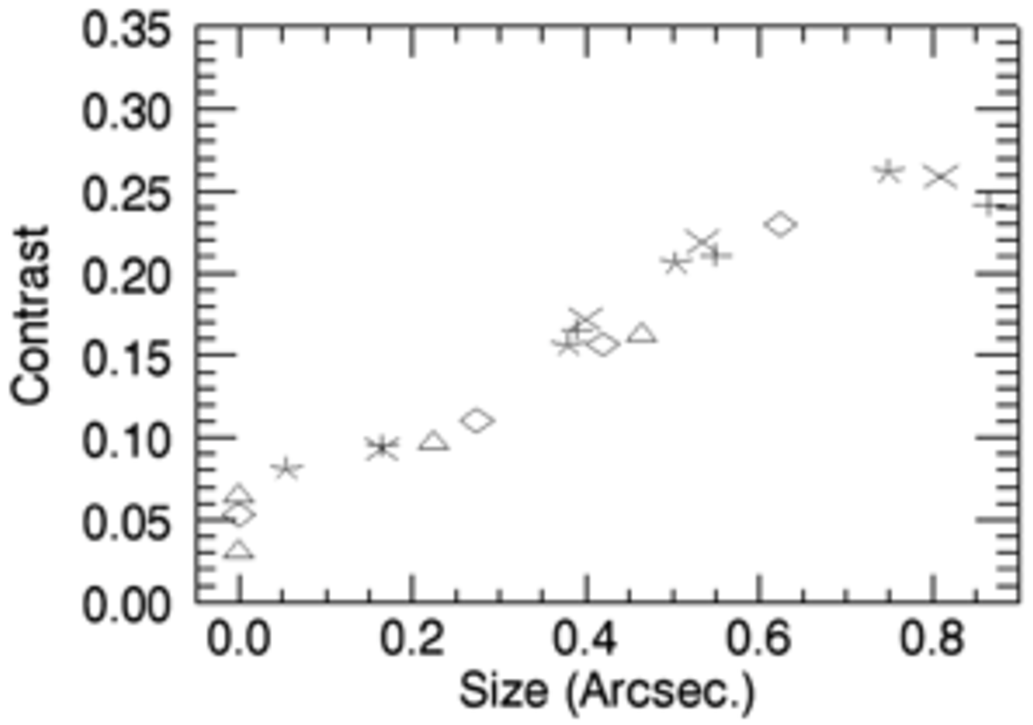}

}
\caption{Size versus contrast of four clusters of flux tubes at various positions on the solar disk. Top: spatial resolution 0.1". Bottom: spatial resolution 0.3".}
\label{CLV_aree_contr}
\end{figure}
In order to investigate whether spatial resolution of observations could explain such discrepancies, we studied the contrast-size relation of various clusters of flux tubes. Figure \ref{CLV_aree_contr} shows the contrast-size relation for an isolated tube
 70 km wide and for clusters of two, three and five tubes 70 km wide and 105 km apart at various positions on the solar disk and for spatial resolution 0.1" (top) and 0.3"(bottom). For vertical directions ($\mu=1$ and $\mu=0.9$) and for resolution 0.1" contrast does not vary significantly with features size. 
At shallower lines of sight, the contrast increases with size. For $\mu=0.8$ and $\mu=0.6$ size dependence of contrast is larger for the smallest sizes while there is no dependence at the largest sizes. 
The increase of the contrast with size at inclined lines of sight is due to the reduced attenuation of radiation in the largest clusters. Moreover, since for each inclination, there is a finite number of flux tubes that radiation can cross (see \S4), there is a size threshold beyond which contrast does not vary with size. This size threshold increases with the decrease of $\mu$.
An increase of contrast with size is observed even at vertical lines of sight with the reduction of spatial resolution, as illustrated in bottom panel of Fig.\ref{CLV_aree_contr}. This is due to the larger reduction in contrast value due to smoothing on the smallest features (see \S4). Bottom plot shows that at each position on the solar disk, data can be well fitted with straight lines, as obtained in observational data by \citet{berger2007}.

\section{Summary and conclusions} 
We have investigated thermal, photometric and geometric  properties of various configurations of clusters of magnetic flux tubes by two dimensional, static numerical simulations. The main results, obtained by the analyses of tubes with Wilson depression 60 km when in RE, are the following.   

At depths for which radiative heating is efficient, the temperature stratification within and around flux tubes is function not only of the physical properties of the flux tubes, but also of the tube number. 
In fact we found that the temperature inside the tubes increases with the number of adjacent tubes in the optically thin part of the domain and decreases at $0<{\rm log}(\tau)<1$. At larger values of optical depth temperature increases again with the filling factor. The area around and between the tubes is cooled at $0.5<\tau<2$, while is heated at greater depths.

Since the temperature stratification determines photometric properties,  contrast profiles of flux tubes also vary in clusters. 
At disk center contrast increases with the number of flux tubes. Moreover regions in between the tubes are characterized by negative contrast, which is a signature of the decrease of temperature, generated by the radiative cooling, in these areas. These dark features gradually disappear with the increase of the inclination of the line of sight. We 
found that for each cluster model there always exist an inclination beyond which clusters appear as compact monolithic features, as if they were isolated tubes. At these inclinations, the contrast of a cluster is larger than the contrast of a single flux tube element, and smaller than the contrast of an isolated tube with the same size and Wilson depression of the cluster. 

As for isolated flux tubes, off-disk center contrast profiles of clusters are very asymmetric, with long tails 
toward the limb and dark lanes at disk center side. Nevertheless, projection onto the plane of the sky and the decrease of the spatial resolution (due for instance to both instrumental and atmospheric degradation), largely makes profiles more symmetric. This finding is in agreement with measurements off-disk center of G-band bright points contrast profiles by \citet{berger2007}. 
These authors found  quite symmetric profiles at all disk positions 
investigated. In both our models and observations, profiles at $\mu=0.6$ are 
more symmetric than profiles at $\mu=0.4$.  This is true in the models
for both isolated and clusters of flux tubes (Fig. \ref{CLV_asym}). Reduction of spatial resolution also decreases the depth and size of dark rings and dark lanes, which disappear in the worst resolution cases investigated (0.3"). Moreover, dark lanes are not ubiquitous features of flux tubes. Since these are signatures of cooler regions within and around the tubes, they are very faint or absent in tubes models with larger radiative heating (see also models presented in \citet{steiner2005} or \citet{pizzo1993b}), or can be observed in some wavelengths and not in others. This helps
explain why \citet{berger2007} did not find dark lanes in all the contrast profiles they observed off-disk center.

CLVs of properties of magnetic bright features are also affected by clustering
and reduction of resolution. The contrast-CLV is for instance steeper for isolated flux tubes than it is
for clusters of tubes. Since most of feature identification techniques on
 images are based on intensity threshold, isolated small (and therefore less brilliant) features are less likely to be detected. Measurements at the limb are therefore biased toward larger tubes and clusters and the resulting CLV is flatter than the one obtained for single structures (see Fig. \ref{CLV_contrast_configs}). For the same reasons, the CLV of sizes is also flatter for tube clusters (see Fig. \ref{CLV_aree_configs}).
 This would explain CLV shapes obtained by recent observations at the extreme limb \citep{hirzberger2005}. Reduction of resolution also shifts the peak of the contrast-CLV toward lower values of $\mu$, 
especially for tubes models with marked dark lanes, and makes the CLVs flatter (see Fig. \ref{CLV_risoluzione}). In this context, we notice that Berger et al. 2007, who analyzed better resolution images with respect to other authors, 
 found the peak of the CLV of contrast to occur closer to the disk center with respect to some previous 
works (see Berger et al. 2007 and references therein).

Finally, we have investigated the size-contrast relation for clusters of various numbers and spatial resolutions. 
We found a slight increase of contrast with size at disk center, which has to be partially ascribed to the increase of temperature with the number of adjacent tubes and partially to the larger effects of smoothing on smaller features. Increase of contrast with size (and therefore with the number of flux tubes) is found for inclined lines of sight due to the reduced opacity inside the tubes.  Since for a given inclination of the line of sight only a finite number of tubes is crossed by radiation (see \S4), there is a size threshold beyond which contrast does not vary. This threshold increases with the inclination of the line of sight. With the decrease of spatial resolution, even for vertical lines of sight the contrast increases with size, the threshold size value increases, and the contrast-size curves obtained at various lines of sight can be fitted by straight parallel lines. These findings are in agreement with recent contrast-size measurements obtained on both high \citep{berger2007} and medium \citep{ermolli2007} resolution images.

Results presented in this work suggest that measurements of properties of magnetic bright features are significantly influenced by spatial resolution, image analysis techniques, and the wavelength of observation. Geometric properties of contrast profiles, such as size and asymmetries, are particularly affected. These findings can partially explain the discrepancies presented in the literature. 
Moreover, some observational results can be better interpreted 
by considering observed bright magnetic features as aggregation 
of smaller elements, rather than a monolithic entity. 

The results presented in this work should be properly taken into account when trying to derive physical properties of magnetic flux tubes by observations.

\begin{acknowledgements}
The authors are grateful to V. Penza for providing the opacity and atmosphere models and to the referee  for the helpful 
review of the manuscript. S.C. acknowledges the OAR grant, which is based on the CVS project supported by Regione Lazio. S.C also thanks I. Ermolli for useful discussions and support.
\end{acknowledgements}

\end{document}